# Colon Polyps Detection from Colonoscopy Images Using Deep Learning

Md Al Amin[a], Bikash Kumar Paul[a,b]

[a] *Department of Software Engineering, Daffodil International University, Daffodil Smart City, Ashulia, Savar, Dhaka 1341, Bangladesh*
[b] *Department of Information and Communication Technology, Mawlana Bhashani Science and Technology University, Santosh, Tangail 1902, Bangladesh*

**Abstract**

Colon polyps are malignant growths that pose a significant risk to both women and men. The most common method for screening these polyps is through colonoscopy-based image analysis. With the recent advancements in machine learning and deep learning, particularly with convolutional neural networks (CNNs), there has been a notable impact on object detection due to their potential for efficient feature extraction. This technology has found extensive use in medical image analysis, particularly in detecting organ abnormalities like colorectal cancer. Given the global prevalence of colorectal cancer, early detection of polyps is crucial for reducing mortality rates. In this study, we utilized a model aimed at early polyp detection. Initially, we collected data from the Kvasir-SEG dataset and applied data augmentation techniques to enhance our dataset's to increase performance. We allocated 80% of the data for training and 20% for testing, with an additional 20% of the training data reserved for validation to prevent overfitting. We employed three versions of the YOLOv5 (You Only Look Once) model – YOLOv5s, YOLOv5m, and YOLOv5l to analyze the dataset and assess their performance. During training, we iteratively increased the number of iterations to improve model performance. Following the final experiment, YOLOv5l demonstrated superior performance compared to YOLOv5s and YOLOv5m. YOLOv5l achieved training accuracy, precision, recall, validation, and training loss, as well as mean average precision (mAP), of 97.00%, 85.00%, 0.0025, 0.035, and 85.1% respectively, affirming its superior detection capabilities. Finally, we evaluated the testing data using all three versions of the YOLOv5 model under various conditions, successfully detecting colon polyp abnormalities and their specific locations. The highest average Intersection over Union (IoU) recorded was 0.86, achieved by the YOLOv5l model.

*Keywords:* Computer vision, Deep learning, YOLOv5, Colon cancer, Colon polys

## 1. Introduction

Cancer is a disease characterized by uncontrolled growth and spread of abnormal cells throughout the body. It is primarily a genetic disorder, often resulting from changes in cellular function caused by genetic mutations [1]. Cancer poses a significant threat globally, with countless individuals affected and succumbing to it each year. Various types of cancer afflict people worldwide, and colorectal cancer is among them. Unfortunately, cancer often lacks a definitive cure, particularly in advanced stages. However, early diagnosis can significantly improve the prognosis and increase the likelihood of successful treatment. Therefore, timely diagnosis and treatment are crucial for managing any type of cancer, including colorectal cancer. With early detection, effective treatment strategies can be implemented to combat colorectal cancer and improve patient outcomes.

The colon, the longest segment of the large intestine, plays a crucial role in the human digestive system. This organ is responsible for processing food, extracting nutrients, and providing fuel for the body [2]. However, like other organs, the colon is susceptible to various issues over time. One common problem encountered is the development of polyps. Polyps are abnormal growths of tissue that can arise within the colon and its lumen. These growths can be classified as either benign or malignant. While benign polyps typically pose minimal risk, malignant polyps represent a precancerous stage [3].In their benign stage, most polyps pose little harm. However, over time, some polyps can progress to colorectal cancer, which becomes highly dangerous if diagnosed in advanced stages [3]. Colorectal cancer can be influenced by genetic or environmental factors, with several risk factors affecting its development, such as age over 50, smoking, overweight or obesity, and lower socioeconomic status [3]. Among all cancer types, colon cancer stands out as one of the most preventable. It ranks as the third most common cancer worldwide [4,5] and is the third most frequently diagnosed cancer in both men and women in the United States [6]. Increasing awareness of health issues has led to more diagnoses occurring at early stages. According to the World Health Organization, in 2018 alone, 1.80 million people worldwide were diagnosed with colorectal cancer, resulting in 862,000 deaths [7]. Annually, nearly 145,600 people are diagnosed with colorectal cancer in the United States, with 101,420 cases attributed to colon cancer and the remainder to rectal cancer [8]. While the mortality rate from colorectal cancer has decreased in recent years, it remains a significant cause of death, with approximately 50,000 deaths reported annually in the USA. Mortality rates have declined slightly among both men and women from 1990 to 2015, with a decrease of 1.8% per year among men and 1.4% per year among women [9]. In Europe, there were an estimated 450,000 new cases of colorectal cancer and 23,200 deaths in 2008, with a mortality rate of 18.86 per 100,000 in 2009 [10]. While colorectal cancer is more common in high-income countries, its incidence is gradually increasing in middle- and low-income countries, albeit at a slower rate in Africa and Asia [11]. Prevailing research consistently underscores the widespread prevalence of colorectal cancer as a global health concern, particularly in middle- and high-income countries. As evidenced by previous studies, while small polyps pose minimal harm when detected early, malignant polyps become significantly harmful when identified at later stages. However, advancements in medical technology enable the early detection and removal of colon polyps, including medium malignant ones. As technology continues to advance rapidly, there is a growing need for high-speed and high-performance capabilities in polyp detection, particularly during colonoscopies. Thus, this paper proposes a model designed to achieve high performance and detect even the smallest objects. We believe this model will significantly benefit the medical sector by addressing the challenge of detecting tiny colon polyps during colonoscopies.

Colorectal cancer (CRC) ranks as the third most common cancer worldwide. Adenomatous polyps, benign growths of glandular tissue, serve as precursors to CRC (adenomas). CRC results from the progression of polyps into malignant tumors over time, often leading to death due to metastasis to various tissues and organs. Real-time colonoscopy, which removes precancerous and small cancerous polyps, offers a higher adenoma detection rate during colonoscopy, thereby facilitating better treatment against CRC. Numerous researchers have explored various methods to enhance the sensitivity and specificity of adenoma detection, aiming to improve the efficiency of medical diagnostic devices and reduce detection errors. This study seeks to employ the most suitable model to enhance sensitivity and specificity during colonoscopy, potentially advancing biomedical engineering or medical sciences. The study progresses as follows: Chapter 2 reviews relevant literature, Chapter 3 discusses the background of object detection and model evolution, and Chapter 4 outlines the methodology. The methodology section covers dataset description, data augmentation techniques, data labeling, model development for colon polyps, and performance metrics analysis. Following the deployment of three versions of the YOLOv5 model, Chapter 4 evaluates each model's performance. The Result and Discussion section comprises evaluations of the training and validation periods, performance and loss matrices, and object detection. Finally, the Conclusions section provides an overview of the research and outlines future prospects.

## 2. Literature Review

This section provides an overview of relevant deep learning (DL) approaches used in the analysis of colon polyps and cancer, as well as augmentation techniques to enhance efficiency.

In Lee et al.'s (2020) study, they developed and tested a deep learning method for polyp detection [12]. Their model, employing the YOLOv2 algorithm, was trained on 8075 images, including 503 polyps. Three datasets were used to validate the model: one comprising 1338 images with 1349 polyps, another from a public CVC clinic with 612 polyps, and videos of colonoscopies containing 26 polyps. The model demonstrated superior sensitivity (96.7%) and specificity (90.2%) with the first two datasets. Interestingly, it detected an additional 7 polyps missed by endoscopists, while achieving an operational speed of 67.16. In Zheng et al.'s (2018) research, they localized polyps using boundary boxes in colonoscopic images employing the YOLO algorithm [13]. Initially trained with non-medical images, the model was further fine-tuned with data collected from different sources. The YOLO model achieved a sensitivity of 68.3% and a precision of 79.3%, with an efficiency of 0.06 seconds per frame. The study suggested that YOLO could effectively assist endoscopists in localizing colorectal polyps during endoscopy.

In Pascal et al.'s (2021) study [14], they developed a model by scaling the YOLOv4 algorithm for real-time polyp detection. To scale YOLOv4, they initially replaced the entire structure with CSPNet and substituted the Leaky ReLU activation function with the Mish activation function. Additionally, they replaced the Complete Intersection over Union (CIoU) with the Distance Intersection over Union (DIoU). Various structures such as Resnet, VGG, Darknet53, and Transformers were utilized to enhance the performance of YOLOv3 and YOLOv4. Their proposed method demonstrated superior performance and accuracy compared to other methods. They achieved precision of 96.04%, recall of 96.68%, and an F1-score of 96.36% on the CVC-ColonDB dataset. In Byrne et al.'s (2019) research [15], they employed an artificial intelligence model with endoscopic video images of colorectal polyps. They utilized a Deep Convolutional Neural Network (DCNN) for training and validating their model. The study involved various series comprising 125 videos with diminutive polyps. The model achieved an accuracy of 94% (95% CI 86% to 97%). In Jha et al.'s (2021) study, they benchmarked several methods for polyp detection, segmentation, and localization accuracy, as well as speed using the Kvasir-SEG open-access colonoscopy image dataset [16]. They demonstrated that most methods achieved a precision of 0.8000, a mean IoU of 0.8100, and a speed of 180 frames per second for detection and localization tasks. Additionally, they reported a coefficient of 0.8206 and an average speed of 182.38 frames per second for the segmentation task. In Yamada et al.'s (2019) research [17], they employed a Faster CNN for detecting early signs of colorectal cancer during colonoscopy. The model exhibited a specificity of 97.3% and a sensitivity of 98.0% in the non-polypoid subgroup. Moreover, the model successfully detected cancerous regions in 21.9 ms/images. This model offers improved accuracy for non-polypoid lesions, which are typically overlooked by optical colonoscopy.

In Wang et al.'s (2018) study, they presented a deep learning model based on the SegNet architecture [18]. They utilized several datasets for training, testing, and validating their model. Their newly collected dataset comprised 27,113 colonoscopy images from 1,138 patients with at least one detected polyp, achieving a sensitivity of 94.38%, specificity of 95.92%, and AUC of 0.984 for validation. In real-time video exploration, the algorithm processed 25 frames per second with a latency of 76.80±5.60 ms when using a multi-threaded processing system. In Yu et al.'s (2016) experiment, they implemented a 3D online and offline integration framework (3D-FCN) for automatic polyp detection [19]. 3D-FCN, possessing greater learning capability representative of spatio-temporal features than 2D-CNNs, combined offline and online learning to limit the number of false positives generated by the offline network and increase detection. Their experiment demonstrated better performance than others with a dataset from the MICCAI 2015 Challenge on Polyp Detection dataset.

In Carneiro et al.'s (2020) exploration, they elucidated the roles of confidence calibration and classification uncertainty in deep learning models [20]. This led them to introduce a new Bayesian deep learning method that enhances classification accuracy and model interpretability by leveraging both calibration and uncertainty. The experiment involved 940 images of colorectal polyps, and the results demonstrate that their proposed strategy achieves state-of-the-art levels of confidence calibration accuracy and classification. In Urban et al.'s (2018) experiments, they devised and trained a deep CNN to detect polyps using 8,641 hand-labeled images from approximately 2,000 patients [21]. They tested their model using 20 colonoscopy videos and validated their work with expert colonoscopists. Their model achieved an area under the receiver operating characteristic curve of 0.991 and an accuracy of 96.4%. In their study, four professional reviewers examined colonoscopy footage in which 28 polyps were removed. They found an additional 8 polyps that had not been detected without CNN assistance and 17 extra polyps that had been removed with CNN assistance (45 in total).

In Appendix A of our research work, we provide a comprehensive overview of our proposed model, detailing its evolution and the enhancements made to improve its efficiency through iterative development.

## 3. Methodology

Colorectal cancer is a highly sensitive disease characterized by the development of polyps in the tissue. Detecting these polyps, especially when they are very small, is crucial for both patients and the medical sector. This study aimed to develop a model specifically designed to detect small polyps effectively. The primary focus of this study was the utilization of the YOLOv5 algorithm, which is based on Convolutional Neural Networks (CNNs). YOLOv5 represents the latest iteration of the YOLO series and is renowned for being the first object recognition method to seamlessly integrate object classification and bounding box prediction within a single, end-to-end differentiable network. Developed and maintained by the Darknet weight, YOLOv5 is notable for being the first YOLO model to be written in the PyTorch framework, making it significantly more user-friendly and lightweight. However, it is worth noting that YOLOv5 does not surpass YOLOv4 in performance on standard benchmark datasets such as COCO. This is primarily due to the lack of fundamental architectural improvements in YOLOv5 compared to its predecessor. Nonetheless, this study employed three versions of the YOLOv5 model - YOLOv5s, YOLOv5m, and YOLOv5 - each of which demonstrated significant efficacy in different scenarios. The workflow of this study is illustrated in Figure 1.

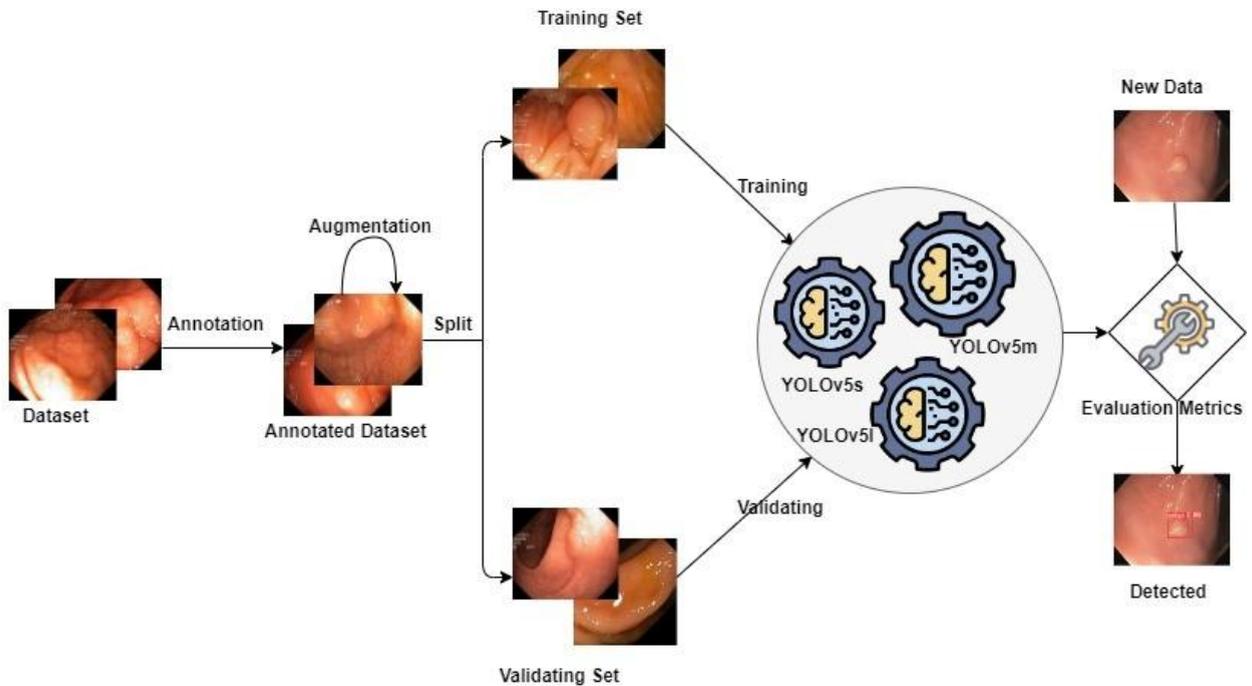

Fig. 1. Colon polyp detection experiment workflow.

### 3.1. Dataset explanation

The proposed object detection method involves training and testing using the Kvasir-SEG dataset [33, 34], which comprises 1000 polyp image data from colonoscopy images. The generated dimensions of the image dataset are 640×640 from various dimensions. However, due to the dataset's limited size, data augmentation techniques were employed to enhance it. Various strategies such as rotation, translation, scaling, horizontal and vertical flipping, etc., were utilized in the augmentation process. As a result of the preprocessing and augmentation techniques, the dataset was expanded, resulting in a total of 1800 images with dimensions of 640×640 pixels. The real dataset is depicted in Figures 2.

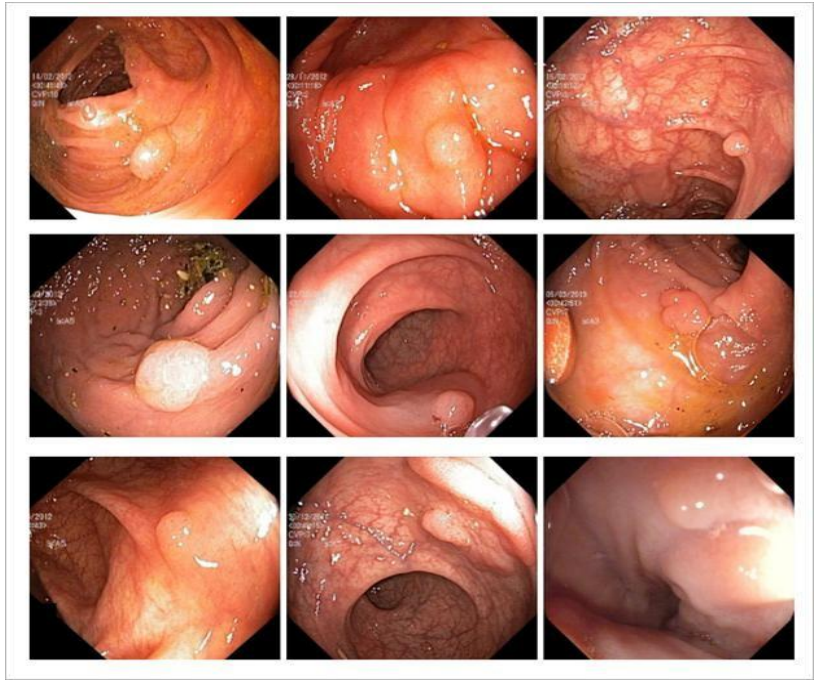

Fig. 2. Colon polyp dataset sample.

*3.2. Data augmentation technique*

We acknowledge that the performance of any model is contingent upon having a robust dataset with a sufficient amount of data. However, our dataset comprises only 1000 datasets, which is not adequate for training our selected model. To address this limitation, deep learning approaches leverage data augmentation techniques to enhance the dataset. Data augmentation involves modifying or combining real data with artificially created data. In this study, image augmentation techniques were employed to create a suitable dataset of polyps for optimal utilization of our model, thereby improving the framework's reliability. Image augmentation has the potential to increase the model's classification and detection accuracy by enhancing the existing picture dataset without the need for collecting new samples or data. Additionally, it significantly diversifies the data available for the model, enriching the dataset for picture categorization from a small image sample dataset. Commonly used data augmentation techniques include adjusting the contrast, hue, brightness, saturation, and geometric transformation, as well as noise adjustment, such as cropping, scaling, and flipping images. For constructing our desired dataset, this experiment employed several crucial data augmentation techniques, including scaling, horizontal flipping, vertical flipping, width and height shifting, rotation, and image reduction. Rotation involved rotating the image clockwise and anticlockwise with angles ranging from 2 to 45 degrees, resulting in the object shifting to various positions. Height and width shifting were applied at 6% and 16%, respectively, to shift the polyp object. Additionally, a scaling process was implemented to reduce the image size by 3% and 16% for augmentation purposes. For image translation, vertical and horizontal shifting ranging from 2% to 9% was used to translate the tumor object to various positions. Furthermore, vertical and horizontal flipping perspectives with probability factors ranging from 0.3 to 0.5 were applied to flip the polyp object to various positions. To illustrate, one image was taken from the real 198 datasets, and five augmentation techniques were applied, resulting in a dataset size of 1800, comprising 720 augmented and 80 randomly generated data points. Finally, after performing the augmentation technique, our randomly augmented techniques are depicted in Figure 3.

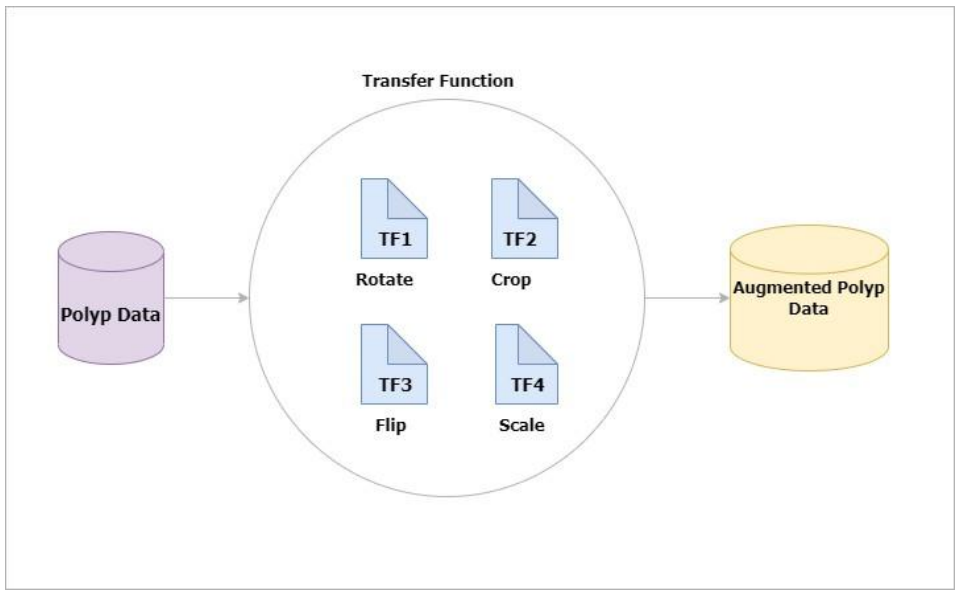

Fig. 3. Data augmentation technique using transfer function (TF)

### 3.3. Data Labeling

Every object detection process requires knowledge of the target's position, which is crucial for detection. The YOLOv5 algorithm utilizes a specific format for detecting the target position, known as labeling. During the labeling phase, files are generated to facilitate the training and validation phases of the YOLO algorithm. The YOLOv5 model is pre-trained using the Microsoft Common Objects in Context (MSCOCO) dataset, which contains 80 predefined object classes. These classes are reserved in the "coco.yaml" file. In our study, we generate another YAML class file named "data.yaml" for labeling YOLO polyp objects. The "data.yaml" file contains only one class, which is the polyp class. The dataset includes polyps of various sizes, categorized as small, medium, etc. Hence, for labeling, we select the polyp class. During the annotation process, each image is annotated in YOLO fashion to generate polyp object labeling files. In the training and validation phases, YOLOv5 utilizes text-format files with the same name as the corresponding images. These files contain the polyps' class ID, center coordinates of the polyp object, and the height and width of the polyp object labeling area. To annotate all polyps training and validation datasets, we utilize the graphical annotation website makesense.ai. Subsequently, the annotated training and validation datasets, along with the corresponding images, are used to train the YOLOv5 network model. As illustrated in Figure 4, the YOLOv5 formatted context file mainly highlights the location of the polyps. Each entry in the text context file represents a target polyp object and includes three elements: the polyp class ID, the bounding box coordinates ($X_0$, $Y_0$), and the area of the polyp (width, height). The ID 0 represents the polyp class.

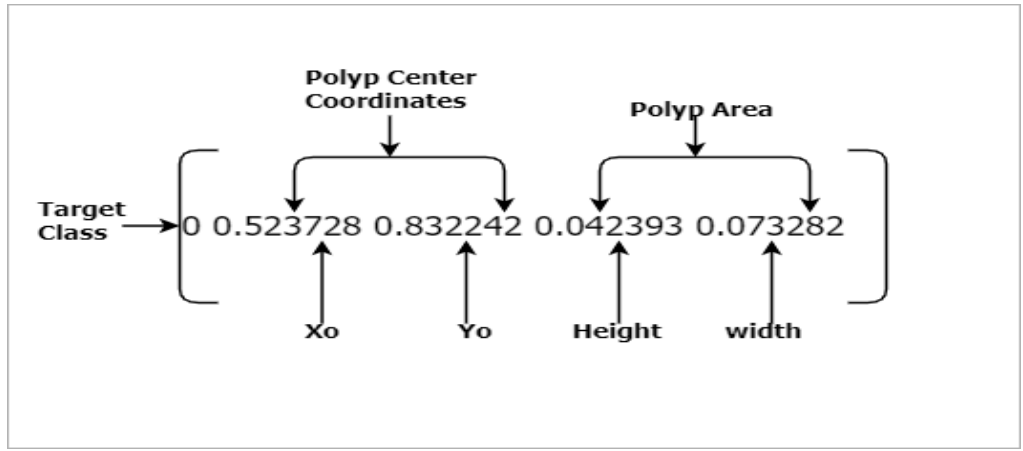

Fig. 4. YOLO format file context for polyp object.

*3.4. YOLOv5 model for polyps detection*

In this study, we employed the YOLOv5 model for its unique advantages. The model offers benefits such as proper object recognition, precise localization of polyps, high speed, and enhanced detection performance. It is capable of detecting small polyp objects in photos that may be noisy, hazy, or cloudy. The YOLOv5 model is a single-stage object detector composed of three important parts: the Backbone, Neck, and Head (Prediction). The base backbone of the YOLOv5 model is the Bottleneck Cross Stage Partial Network (BCSPN), which takes input images with dimensions of $640 \times 640 \times 3$ pixels. The input image passes through the backbone's FOCUS module, which splits the image into four smaller parts for convolutional operations. After 32 convolutional operations, the image becomes a $320 \times 320 \times 3$ feature pyramid. The BCSP module pulls features from the feature pyramid and incorporates critical information from the training images. It eliminates gradient information duplication in the CNN's optimization process and reduces input parameters and model size. The BCSP module consists of two CoBL modules and an adder contained in a residual unit. The adder accumulates the features of the preceding CoBL module outcome and the features of the two CoBL modules, then transfers the endemic characteristics to $1 \times 1$ Conv2D layers. YOLOv5 offers four different models: YOLOv5s, YOLOv5m, YOLOv5l, and YOLOv5x. Each model adjusts the BCSP module with width (w) and depth (d) parameters. Additionally, the BCSP module connects to the spatial pyramid pooling (SPP) in the backbone, which increases the network's receptive field and adds scale-specific functionality. The neck part of YOLOv5 integrates the Path Aggregation Network (PANet) to improve information flow[27]. PANet utilizes a feature pyramid network (FPN) to transfer robust semantic characteristics from top to bottom and expresses strong positional characteristics from bottom to top. This enhances the propagation of low-level characteristics and the use of precise localization signals, thereby improving the accuracy of target object positioning. The head layer, also known as the prediction or detection layer, constructs three separate feature pyramids for multi-scale prediction. In the detection layer, the model can identify and recognize objects of varying sizes, including tiny, medium, and large objects,

Step 1: Initially, the backbone receives images with a resolution of $640 \times 640$ pixels. Then, the FOCUS module splits the images. After executing multiple convolutions and two BCSP1 operations, the feature pyramid is transferred to the second concatenation layer. In contrast, the feature pyramid travels to the second concatenation layer after completing BCSP1, two BCSP2 operations, a set of convolution operations, and two upsampling operations. Both of these paths are concatenated in the second concatenation layer. After completing the BCSP2 layers and $1 \times 1$ convolution operation, the feature pyramid, sized $80 \times 80$, is obtained as scale 1.

Step 2: In this step, a $3 \times 3$ convolutional kernel processes the $80 \times 80$ sized feature pyramid from step 1 and sends it to the third concatenation layer. Additionally, a $1 \times 1$ convolutional kernel processes the feature pyramid before the second upsampling and sends it to the third concatenation layer. Both of these outputs are concatenated in the third concatenation layer. After completing the BCSP2 layers and the $1 \times 1$ convolution operation, the $40 \times 40$ sized feature pyramid, scaled 2, is obtained.

Step 3: In this step, a $3 \times 3$ convolutional kernel processes the $40 \times 40$ sized feature pyramid from step 2 and sends it to the fourth concatenation layer. Additionally, a $1 \times 1$ convolutional kernel processes the feature pyramid before the second upsampling and sends it to the fourth concatenation layer. Both of these outputs are concatenated in the fourth concatenation layer. After completing the BCSP2 layers and $1 \times 1$ convolution operations, the $20 \times 20$ sized feature pyramid, scaled 3, is obtained.

Step 4: Finally, we obtain different shapes of feature pyramids on a scale of 1 to 3 (i.e., 80×80, 40×40, 20×20). These improved feature pyramids are responsible for detecting various-shaped polyp objects with regression bounding boxes. Therefore, three regression bounding boxes are predicted for each feature pyramid at every location. Consequently, a total of 25,200 regression bounding boxes are created ($3 \times 80 \times 80 + 3 \times 40 \times 40 + 3 \times 20 \times 20$). The final polyp detection method with the bounding objects is illustrated in Figure 5.

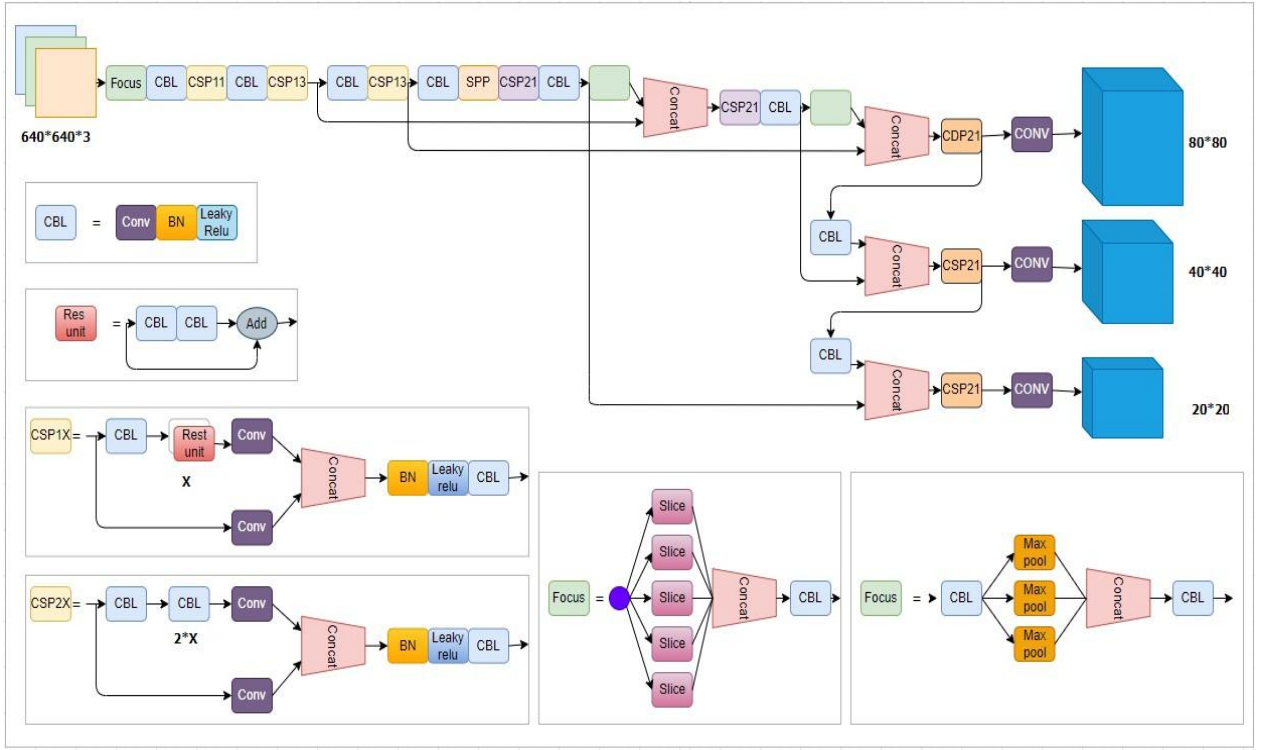

Fig. 5. Network Diagram of YOLOv5 model

*3.5. Environment setup and training explanation*

This experiment was primarily conducted on the Google Colab Platform, utilizing an NVIDIA graphics card and a high-power CPU processor with ample memory capacity. The model was trained using the PyTorch deep learning framework and Python version 3.8. Three different versions, namely YOLOv5s, YOLOv5m, and YOLOv5l, were employed in this experiment to compare their performance on the polyp tumor dataset. The experiment utilized the final training dataset, which was annotated with class files for training the desired model. After augmentation, the dataset comprised 800 images, including real data, along with annotated class data for all images. The training phase utilized 80% of the data for training and 20% for testing, with the training data further split into two parts for training and validation to prevent model overfitting. The modified data.yaml file and pre-trained weights were employed during the training period for all models. The model was configured with a learning rate of 0.01, a batch size of 16, and the default optimizer stochastic gradient descent (SGD) to optimize model performance. A summary of the model hyperparameters is presented in Table 1.

Table 1. YOLOv5 model training hyperparameter

| Parameter's name | Value |
|---|---|
| Initial Learning Rate | 0.01 |
| Momentum | 0.937 |
| Opimizer | SGD |
| Learning Factor | 0.2 |
| IoU Training Thresold | 0.20 |
| Anchors | 3.0 |
| Weight Decay | 0.0005 |
| Warmup Epochs | 3.0 |
| Warmup Momentum | 0.8 |
| Initial Warmup Bias Learning Rate | 0.1 |
| Box loss gain | 0.05 |
| Anchor Multiple Thresholds | 4.0 |

*3.6. Performance evaluation metrics*

Evaluate object detection using several performances matrices like Precision, Recall, Average Precision (AP), Mean average precision (mAP), Intersection over union(IoU), and Generalized intersection over union(GIoU) as instances of the most utilized criteria [29].

In straightway, Precision is the ratio against the true positive rate and all the true positive rates. In our cases, it would be the number polyp image that the model correctly identifies out of all the polyp consisting images. Precision value define as follows,

$$\Pr ecision(P) = \frac{TruePositive(TP)}{TruePositive(TP) + FalsePositive(FP)}$$
(1)

In the simplest way, Recall is a measure of how well our model detects True Positives. As a result, recall tells us how many patients we accurately identified as having colon polyps out of all those who have it. Recall value define as follows,

$$\operatorname{Re} cal(R) = \frac{TruePositive(TP)}{TrueNegative(TN) + FalseNegative(FN)}$$
(2)

Average Precision (AP) is another way to evaluate the object detection framework. Calculate the average precision using precision and recall. It summarize the Precision-Recall curve using average the recall digit 0 to 1. It is mainly an individual number metric [30]. In mathematically it used eleven-point interpolated average precision.

$$Average\,Precision = \frac{1}{2}\sum_{r\in(0,0.1,0.2....1)} Pnterp(r)_i \qquad (3)$$

Mean Average Precision (MAP) is a way if the dataset has M class then calculating mAP. It generally accepts AP average over M classes [31]. MAP value define as follows,

$$MeanAverage\,Precision(mAP) = \frac{1}{M}\sum_{j=1}^{M} AP_j \qquad (4)$$

The bounding box loss regression define as follows

$$l_{box} = \lambda_{crd}\sum_{i=0}^{S^2}\sum_{j=0}^{B} I_{i,j}^{obj} bj(2 - w_i \times h_i)\left[(x_i - \hat{x}_i^j)^2 + (y_i - \hat{y}_i^j)^2 + (w_i - \hat{w}_i^j)^2 + (h_i - \hat{h}_i^j)^2\right] \qquad (5)$$

Where $\lambda_{crd}$, loss coefficient position, $\hat{x}, \hat{y}$ is the actual central coordinate of the target and $\hat{w}, \hat{h}$ is actual height and width of the target. And $I_{i,j}^{obj}$ is target anchor box where $(i, j)$ is the target position. $I_{i,j}^{obj}$ value is 1, if the anchor box $(i, j)$ contains target otherwise value is 0.

The confidence loss function define as follows

$$l_{obj} = \lambda_{noobj}\sum_{i=0}^{S^2}\sum_{j=0}^{B} I_{i,j}^{noobj}(c^i - \hat{c}_l)^2 + \lambda_{obj}\sum_{i=0}^{s^2}\sum_{j=0}^{B} I_{i,j}^{obj}(c^i - \hat{c}_l)^2 \qquad (6)$$

Where $\lambda_{noobj}$, number of object loss, $c_i$ is predicted confidence and $\hat{c}_l$ is the actual confidence. And $I_{i,j}^{noobj}$ is number of target anchor box where $(i, j)$ is number of target object position. Here, $\lambda_{obj}$ is target object, and $I_{i,j}^{obj}$ is target anchor box where $(i, j)$ is target object position.

Intersection over Union (IoU) generally discovers the difference between the ground truth and the detected bounding boxes. In the state-of-the-art object identification algorithm, the assessment criteria are applied. During the object detection phase, the algorithm predicts numerous bounding boxes for each object and eliminates groundless boxes based on the confidence point of each bounding box's threshold standard corner stone. Ordinarily, Based on the required threshold value is determined. If the threshold value is lower than the *IoU* value then take the object, otherwise, the bounding box eliminates [32]. *IoU* value define as follows,

$$Intersection\,Over\,Union(IoU) = \frac{A \cup B}{A \cap B} \qquad (7)$$

Where *A* and *B* are two arbitrary convex shapes that perform union and intersection operations then divide the area of the overlap by the area of union.

## 4. Result and discussion

Upon completion of all experiments, this section delves into the performance of the experiments and compares the effectiveness of the YOLOv5 model. The model, trained using the latest YOLO series, demonstrates high proficiency in detecting small objects. In our experiment, we utilized an 1800-polyp image dataset, allocating 80% of the data for training and 20% for validation. In real-time medical settings, machines sometimes overlook very small polyps, posing significant risks to patients. However, the YOLOv5-based model exhibited enhanced performance in detecting small polyps.

### 4.1. Training and validation evaluation

We utilized several versions of YOLO for the same dataset, namely YOLOv5s, YOLOv5m, and YOLOv5l. Each model exhibited varying performances during the training and validation phases. To enhance model performance and minimize training and validation loss, we conducted multiple epochs. Following all operations on our selected three models, we observed that YOLOv5l outperformed the other two models, with significantly lower validation and training loss. We conducted training and validation for all selected models using 50, 70, and 100 epochs. Initially, after 50 epochs, all models demonstrated lower accuracy. Subsequently, after 70 epochs, the performance of all selected models improved compared to the previous training phase. Finally, after 100 epochs, our model's training and validation performance became more viable, achieving better results than the other two models. The investigation results of our model's training bounding box losses, training object losses, validation bounding box losses, and validation object losses are depicted in figures 6, 7, and 8.

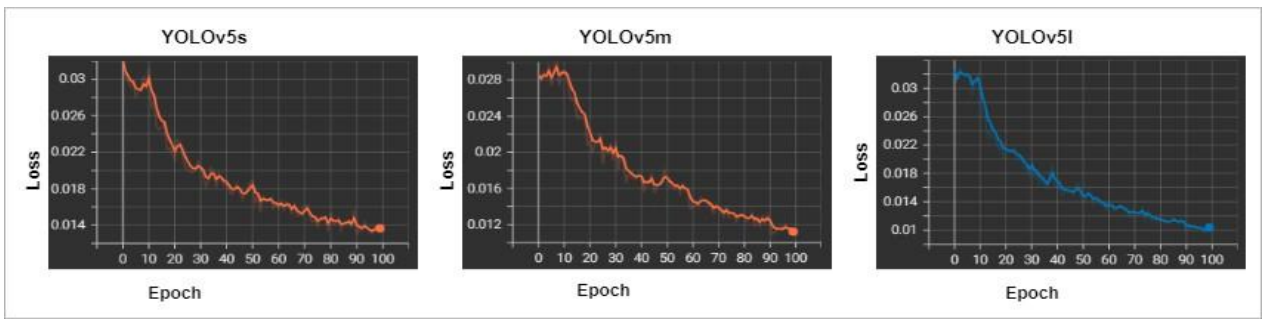

Fig. 6. Training object losses

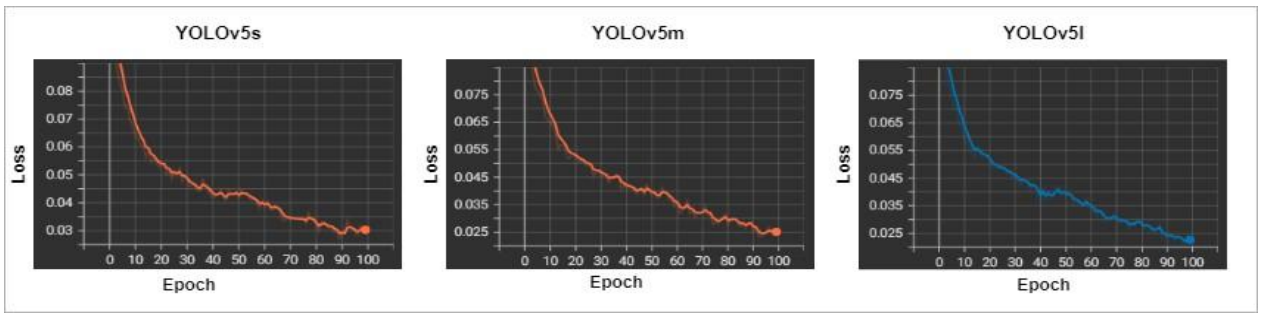

Fig. 7. Validating bounding box losses.

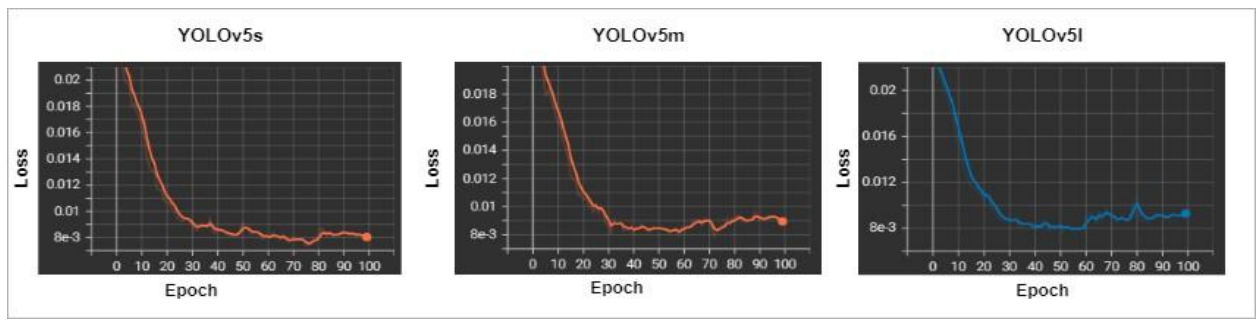

Fig. 8. Validating object losses

## 4.2. Performance and loss metric evaluation

We investigated our employed model using only one class across three versions of the YOLOv5 framework: YOLOv5s, YOLOv5m, and YOLOv5l. We compared the performance of these versions based on evaluation metrics such as precision, recall, average precision, mean average precision, training loss, and validation loss. During the initial phase with 50 epochs, the precision rate varied between 75% to 85% for YOLOv5s, 75% to 86% for YOLOv5m, and 77% to 85% for YOLOv5l. With 70 epochs, there was an increase in the precision rate across all models. Finally, after 100 epochs, YOLOv5s achieved a precision rate of 90% to 95%, YOLOv5m achieved 95% to 97%, and YOLOv5l achieved 92% to 97%. Similarly, the recall rates were initially low for all models but improved in the second phase of the experiment. In the final phase, recall rates ranged between 80% to 85% for YOLOv5s, 75% to 85% for YOLOv5m, and 80% to 85% for YOLOv5l. Regarding other evaluation metrics, such as average precision and mean average precision, the pattern remained consistent across all phases, with improvements observed as the experiment progressed. In terms of training and validation loss metrics, YOLOv5s exhibited losses ranging from 7% to 8%, YOLOv5m from 7.5% to 1%, and YOLOv5l from 7.5% to 1%. After analyzing all the evaluation metrics, it is evident that the YOLOv5l model consistently demonstrated the best performance among the three frameworks. The performance metrics evaluation is illustrated in figures 9, 10, 11, and 12.

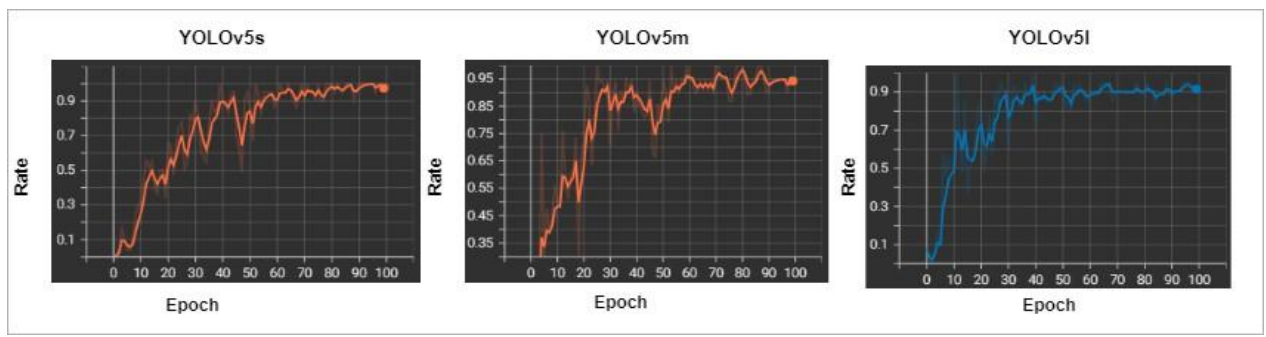

Fig. 9. Precision compare three model.

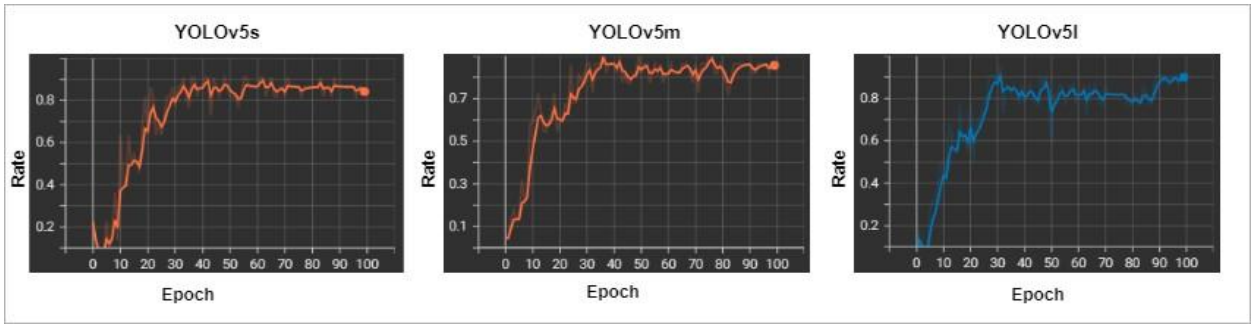

Fig. 10. Recall compare three employed three model

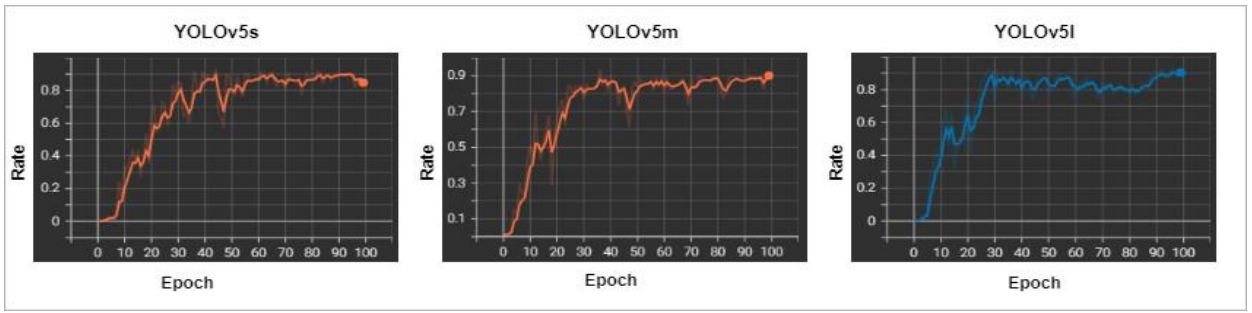

Fig. 11. Mean average precision using IOU threshold at 0.5 compare three employed model

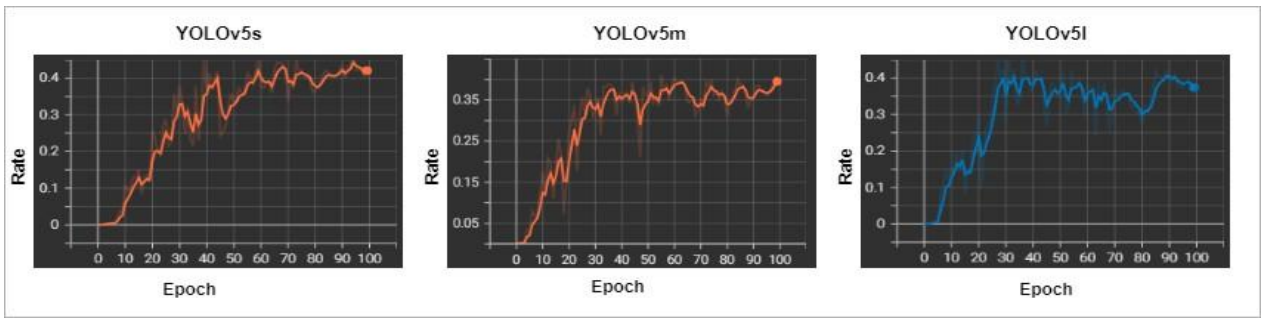

Fig. 12. Mean average precision using IoU thresholds at 0.5 and 0.95. and compare three employed model.

### 4.3. Object detection result evaluation

In In the object detection process, various benchmarking criteria are utilized to assess the performance of models. One such criterion is the Intersection over Union (IoU) method, which evaluates detection performance by calculating the overlap between predicted bounding boxes and ground truth bounding boxes. During our testing phase, we evaluated the IoU scores of different models using a variety of images. The YOLOv5s model achieved IoU scores ranging from 70% to 80%, the YOLOv5m model ranged between 75% to 85%, and the YOLOv5l model ranged from 80% to 90%. Overall, the YOLOv5l model consistently demonstrated the highest IoU scores, indicating better detection performance compared to the other models. Figure 13 illustrates the testing results for the YOLOv5s model, Figure 14 for the YOLOv5m model, and Figure 15 for the YOLOv5l model. Additionally, Table 2 provides a summary of the performance of these models on various testing images.

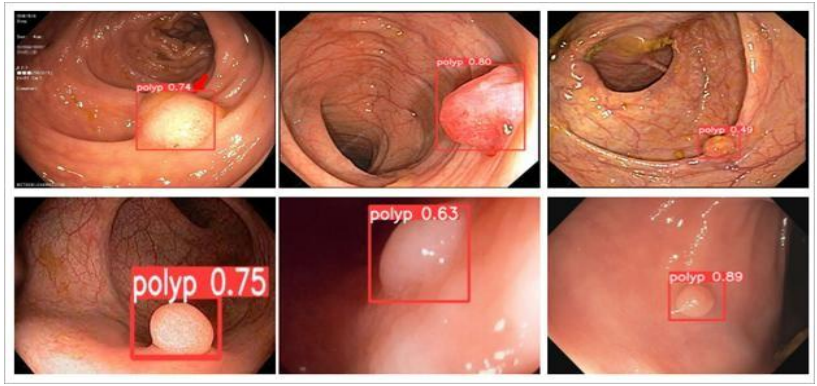

Fig. 13. YOLOv5s model polyp detection result

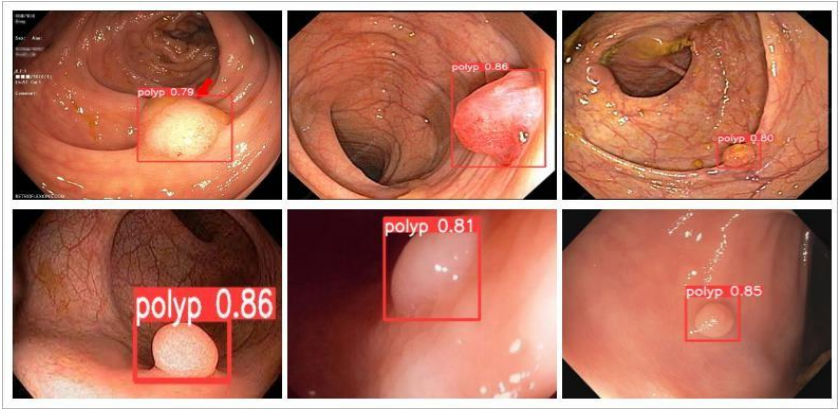

Fig. 14. YOLOv5m model polyp detection result.

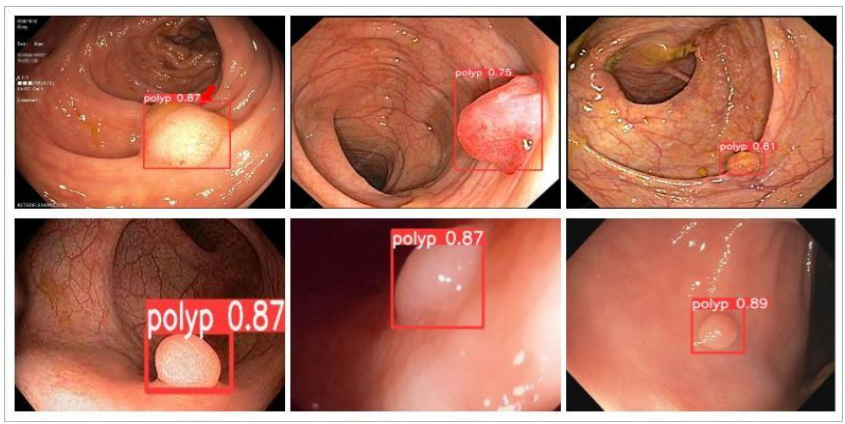

Fig. 15. YOLOv5l model polyp detection result

Table 2. Average IoU for all testing image for each model.

| Model | Average Result(IOU) |
|---|---|
| YOLOv5s | 0.71 |
| YOLOv5m | 0.82 |
| YOLOv5l | 0.86 |

**5. Conclusion**

This study begins by exploring the background of object detection technology and highlights its potential to address various challenges in the medical sector. Following an analysis of our dataset, we applied several data augmentation techniques to enhance our dataset and improve model performance. The final dataset comprised 1800 images, including original images, which were utilized for training, validation, and testing the YOLOv5 model. For the experiment, 80% of the data was allocated for model training, while the remaining 20% was used for validation. The YOLOv5 framework employed CoBL modules, BCSP module, Leaky ReLU activation function, and multiple 3×3 and 1×1 convolutional layers to optimize model features and create a feature pyramid across three scales. Upon conducting the experiment, YOLOv5l outperformed the YOLOv5s and YOLOv5m models in polyp object detection. The YOLOv5l framework exhibited high precision, recall, training accuracy, F1 score, and mean average precision (mAP), along with low training and validation loss. Additionally, during the testing phase, the YOLOv5l model demonstrated superior performance compared to the other models, accurately detecting polyp objects and providing appropriate detection locations with detection scores. Our implementation of the YOLOv5 model demonstrates its efficacy in small object detection tasks, particularly in detecting small colon polyps from colonoscopy images. This advancement holds promise for revolutionizing the medical sector and represents a significant step forward in medical image analysis.

**Acknowledgements**

Acknowledgements and Reference heading should be left justified, bold, with the first letter capitalized but have no numbers. Text below continues as normal.

**Appendix A. YOLO model analysis**

Object detection plays a vital role in modern society across various sectors such as security, military, medical, traffic management, and everyday life, aiming to enhance convenience, health, and safety. With the rapid advancement of deep learning technology, particularly Convolutional Neural Networks (CNNs), object detection models have significantly evolved. Currently, two main techniques are employed for object detection: single-stage algorithms, represented by YOLO (You Only Look Once), and two-stage algorithms, represented by RCNN (Region-based Convolutional Neural Network). While single-stage methods typically exhibit faster detection speeds than two-stage algorithms, they may have lower detection accuracy. Consequently, single-stage algorithms like YOLO are preferred for real-time object detection tasks due to their lightweight nature and high detection speed. The YOLO series, first introduced by Redmon et al. [22] in 2016, has seen several iterations to enhance model performance. YOLOv1 was the inaugural version, which generated candidate regions before object detection. Despite its relatively high detection rate, YOLOv1 suffered from slow processing speeds. YOLOv2, an improved version, utilized a multi-scaled feature pyramid based on Single Shot Multibox Detector (SSD) principles. It incorporated anchor mechanisms and batch normalization techniques to mitigate overfitting and expedite convergence. YOLOv3 further refined the architecture by integrating residual models for network depth and Feature Pyramid Networks (FPN) for multi-scale detection. Darknet-53, based on ResNet, was used as the backbone network for feature extraction, and the model employed a regression-based frame prediction method for object localization. YOLOv4, proposed by Alexey et al. [25], aimed to enhance speed and accuracy. It introduced CSPDarknet for feature extraction, which improved accuracy while reducing computational overhead. YOLOv4 also leveraged the Mish activation function and cross-stage hierarchy combination to optimize target detection settings. Additionally, YOLOv4 utilized the PANet for information circulation and fused resolution feature pyramids for object detection across multiple layers. YOLOv5, developed by Ultralytics [26], continued to innovate by utilizing CSPDarknet for feature extraction and BottleneckCSP for feature pyramid generation. The Fcos algorithm was introduced to enhance frame selection area calculation, thereby improving detection efficiency. Advanced data augmentation techniques were employed to maximize dataset utilization, resulting in breakthrough performance in target detection frameworks. In summary, the evolution of YOLO models has led to significant advancements in object detection, with each iteration introducing novel techniques to improve speed, accuracy, and efficiency in various applications.